\documentclass[preprint]{elsarticle}
\usepackage{geometry}
\usepackage{graphicx}
\usepackage{subcaption}
\usepackage{float}
\usepackage{amsmath}
\usepackage{pdfpages}
\usepackage{hyperref}

\journal{International Journal of Modern Physics C}

\begin{document}
\begin{frontmatter}

        \title{GPU Acceleration of Swendson-Wang Dynamics}
        \author[a]{Tristan Protzman\corref{cor1}}
        \author[b]{Joel Giedt\corref{cor2}}
        \address[a]{Department of Physics, Lehigh University, Bethlehem, PA 18015}
        \address[b]{Department of Physics, Applied Physics, and Astronomy\\Rensselaer Polytechnic Institute, Troy, NY 12180 USA}

        \cortext[cor1]{\texttt{tlp220@lehigh.edu}}
        \cortext[cor2]{\texttt{giedtj@rpi.edu}}

        \date{\today}

    \begin{abstract}
	When simulating a lattice system near its critical temperature, local algorithms for modeling the system's evolution can introduce very large autocorrelation times into sampled data.  This critical slowing down places restrictions on the analysis that can be completed in a timely manner of the behavior of systems around the critical point. Because it is often desirable to study such systems around this point, a new algorithm must be introduced.  Therefore, we turn to cluster algorithms, such as the Swendsen-Wang algorithm and the Wolff clustering algorithm.  They incorporate global updates which generate new lattice configurations with little correlation to previous states, even near the critical point.  We look to accelerate the rate at which these algorithm are capable of running by implementing and benchmarking a parallel implementation of each algorithm designed to run on GPUs under NVIDIA's CUDA framework.  A 17 and 90 fold increase in the computational rate was respectively experienced when measured against the equivalent algorithm implemented in serial code.  
    \end{abstract}
    \end{frontmatter}

    \section{Introduction}

         Around the critical point(s) of a system modeled on a lattice, many algorithms experience the phenomenon of critical slowing down.  This occurs when the critical exponent of the algorithm in use is sufficiently large that data collected around the point of a phase transition has large autocorrelation times.  The behavior can be modeled as $\tau \sim |g-g_*|^{-z}$, where $\tau$ is the autocorrelation time, $g$ is the coupling or temperature and $g_*$ is its critical value.  $z$ is the critical exponent which controls the slow-down; for a diffusion based algorithm, such as a local Metropolis-Hastings update, one would have $z=2$.  This limits our ability to study systems around the point $g_*$ as it greatly increases the number of iterations needed to collect statistically uncorrelated results.  However, there exist clever algorithms which possess a very small critical exponent $z \approx 0$.  One such process is known as Swendsen-Wang dynamics \cite{rcm1}.

         Swendsen-Wang dynamics is able to offer small autocorrelation times because of how it generates new lattice configurations.  Instead of attempting a localized change to the lattice and conditionally accepting it at each iteration as in the Metropolis-Hasting algorithm, clusters of sites (often large) on the lattice are considered and potentially changed as a whole at each iteration.  This allows for subsequent configurations to be significantly different because the modifications are non-local.  Crucially, it is not a diffusive process.  Therefore, large, collective fluctuations can be simulated efficiently so that the system can be better studied around the critical point.
         
         However, because it is advantageous for scaling studies to simulate as large of a system as possible, it is still important to take advantage of the parallel computing capabilities of modern computers.  Note that virtually all performance gains since around 2002 have come from increasing the number of computing cores on processors or the use of accelerators such as graphical processing units (GPUs).  So, in order to realize the advances of the last $\sim$20 years, it is essential to have a parallel algorithm.  The sites of the lattice provide a natural division of work; i.e., a unique thread associated with each site.  The limiting factors of such a scheme are memory bandwidth and latency, not computational cost.  GPUs have a large memory bandwidth, but high latency on the main memory accesses.  Therefore they have both opportunities and challenges; a successful strategy requires that the low latency registers, caches and shared memory be used in a maximal way and that main memory accesses be coordinated to hide latency by having large data streams in flight.  This heterogeneity in the memory hierarchy as well as the locality of caches and shared memory to streaming multiprocessors introduces a complexity that the programmer must tackle in order to make efficient use of the device.  In this article we illustrate an approach that addresses these issues.

    \section{Swendsen-Wang Algorithm} 
        The Swendsen-Wang algorithm is an iterative process to generate a new configuration of spins consistent with the partition function of the system,
        \begin{equation}
            Z = \sum_{\{\sigma\}} e^{-H(\{\sigma\})/k_B T}
        \end{equation}  
        There are five steps to the algorithm, of which three require computation.  Here each step and its implementation will be described.  In what follows, we will be discussing the two-dimensional Ising system on a square lattice.  The approach can be generalized to other discrete spin systems on other lattices.

        \subsection{Bond Formation}

            The algorithm starts with the formation of bonds between sites with like spins.  Starting from an initial configuration (Fig.~\ref{fig:initialConfig}), each site attempts to form a bond with neighboring sites in the North, East, South, and West directions.  In addition to the requirement of sites having equal spins, a probability of $p=1-\exp(-1/k_B T)$  governs the formation of bonds.  This has the effect of forming large clusters of bonded sites when the temperature is low and smaller clusters for higher temperatures.  A depiction of bonds formed is given in Fig.~\ref{fig:formingBonds}.

            Achieving this in parallel is trivial.  Every site is assigned its own thread which checks the sites to East and South for its spin.  Given a match, a random number is generated and used to determine whether or not a bond is formed, with the above-mentioned probability.  To increase the speed of checking locations, shared memory is utilized, loading neighboring tiles into it.  This is worthwhile for two reasons: the first is that each site is checked twice; therefore we reduce our global memory accesses by a factor of two.  Secondly, by putting all the data loads early in the kernel's execution, it allows for the kernel to be completed in a coalesced fashion before any threads diverge.\footnote{A segment of code that executes on the GPU is commonly called a ``kernel.''  Threads ``diverge'' when conditional statements cause them to execute different logical paths.}


        \subsection{Erasure of Spins}
            Once bonds have been formed between the appropriate sites, all current spins can be disregarded.  Since the new configuration is based on the clusters of sites and not the prior spins, they are irrelevant.  While this step is dictated by the algorithm, it requires no attention in code as the newly generated spins will overwrite existing values.  

        \subsection{Cluster Formation}
            Once bonds exist where appropriate, the lattice must be divided into the clusters of connected sites.  These clusters contain all the sites which should be assigned the same spin value in the following step.  Since this is a nonlocal search, completing this step in parallel was the most challenging to implement properly and efficiently.  The algorithm selected to complete this task is a ``label equivalence algorithm'' described by Kalentev et.~al \cite{clustering1}.  Additional descriptions of the algorithm are given in \cite{clustering2,clustering3}.  The algorithm works in a three step iterative process where each pass refines the clustering until it is complete.  A depiction of complete clustering is shown in figure \ref{fig:formingClusters}.

            In the initialization of the algorithm, each site is labeled with a unique integer incrementing from zero.  This initial label serves as the basis for clustering, as each site attempts to find the lowest value label it is connected to.  It does so by creating a forest of references.  Each site does so by comparing its label to those of its neighbors and forming a reference to the lowest valued neighbor.  These forests are then collapsed into trees by a function which follows and links each site to the lowest label in its chain of references.  It then looks up the label it has been decided it is equivalent to, and assigns it to itself.

            This process will repeat until every site is divided into its appropriate cluster.  For lattices sized $512^2$, this process generally completes in 5 or 6 parallel iterations.

        \subsection{Assignment of New Spins}
            After all clusters are formed, each cluster is assigned a new random spin value, with an equal probability of it being up or down.  All sites are then assigned the spin of the cluster it is contained within.  This creates a new lattice configuration which can be very different than the prior one while still preserving the expected characteristics of the system. Since these are large moves that affect many spins in the case of large clusters, the algorithm has much better behavior in sampling configuration space than diffusive algorithms such as Metropolis.  Figure \ref{fig:newConfig} depicts the new configuration generated by this algorithm.

        \subsection{Erasure of Bonds}
            We can now disregard the bonds formed between sites, leaving a new configuration of spins with little correlation to the prior configuration.  \\

        For each new configuration, the average magnetization and energy of the lattice is measured and recorded.  Average magnetization is given by $M=\frac{1}{N}\sum^N_{i=1}S_i$, while the energy is given by the Hamiltonian $H=-\sum_{\langle i,j\rangle}\sigma_i\sigma_j$.  These sums are efficiently calculated by reduction routines provided by the CUB library \cite{cubLibrary}.  

		\newcommand{\figureWidth}{0.42\linewidth}
        \begin{figure}[H]
            \centering
            \begin{minipage}{\figureWidth}
                \centering
                \textbf{Initial Configuration}\par\medskip
                \includegraphics[width=0.8\linewidth]{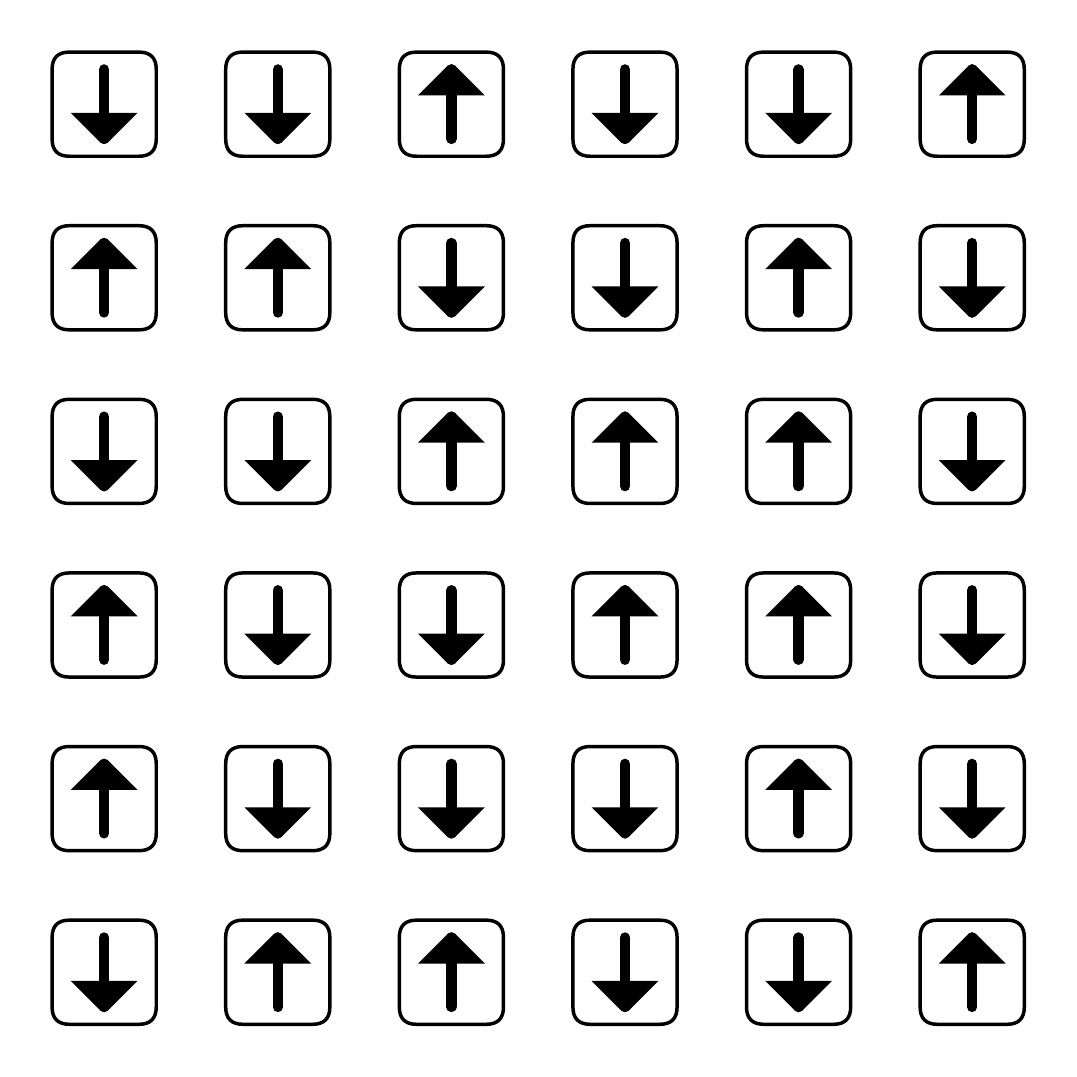}
                \caption{The algorithm starts with an initial configuration of spins.  For the first iteration, this is generated by randomly assigning a spin to each site.  For all other iterations, it is the state the system is in at the completion of the prior iteration.}
                \label{fig:initialConfig}
            \end{minipage} \qquad
            \begin{minipage}{\figureWidth}
                \centering
                \textbf{Formation of Bonds}\par\medskip
                \includegraphics[width=0.8\linewidth]{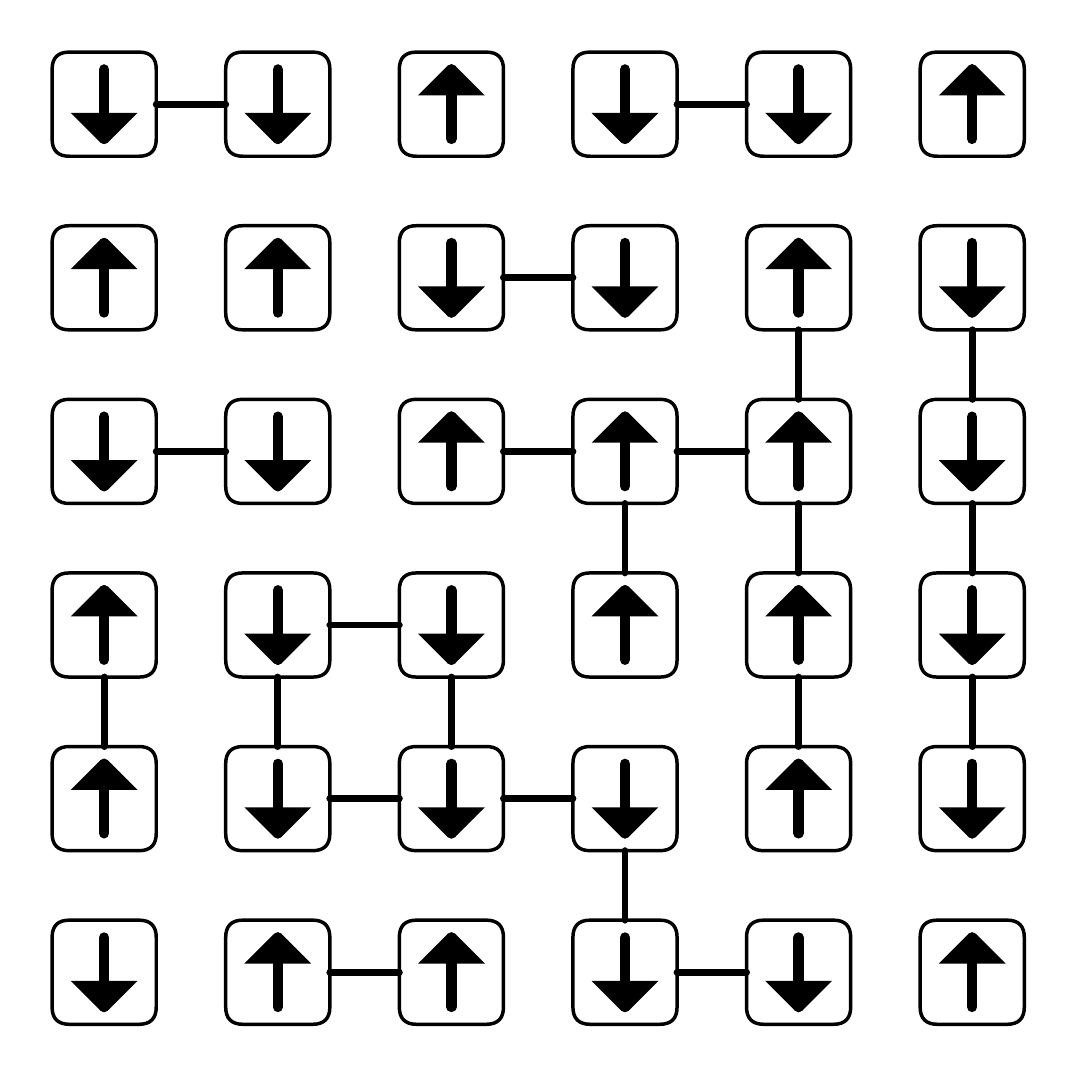}
                \caption{Bonds are formed between neighboring sites if they have the same spin and the temperature allows.  This is aided by the GPU's many threads.  Each site is assigned a thread which checks the neighbor below and to the right to determine if a bond should be formed.\\}
                \label{fig:formingBonds}
            \end{minipage} \qquad

            \begin{minipage}{\figureWidth}
                \centering
                \textbf{Formation of Clusters}\par\medskip
                \includegraphics[width=0.8\linewidth]{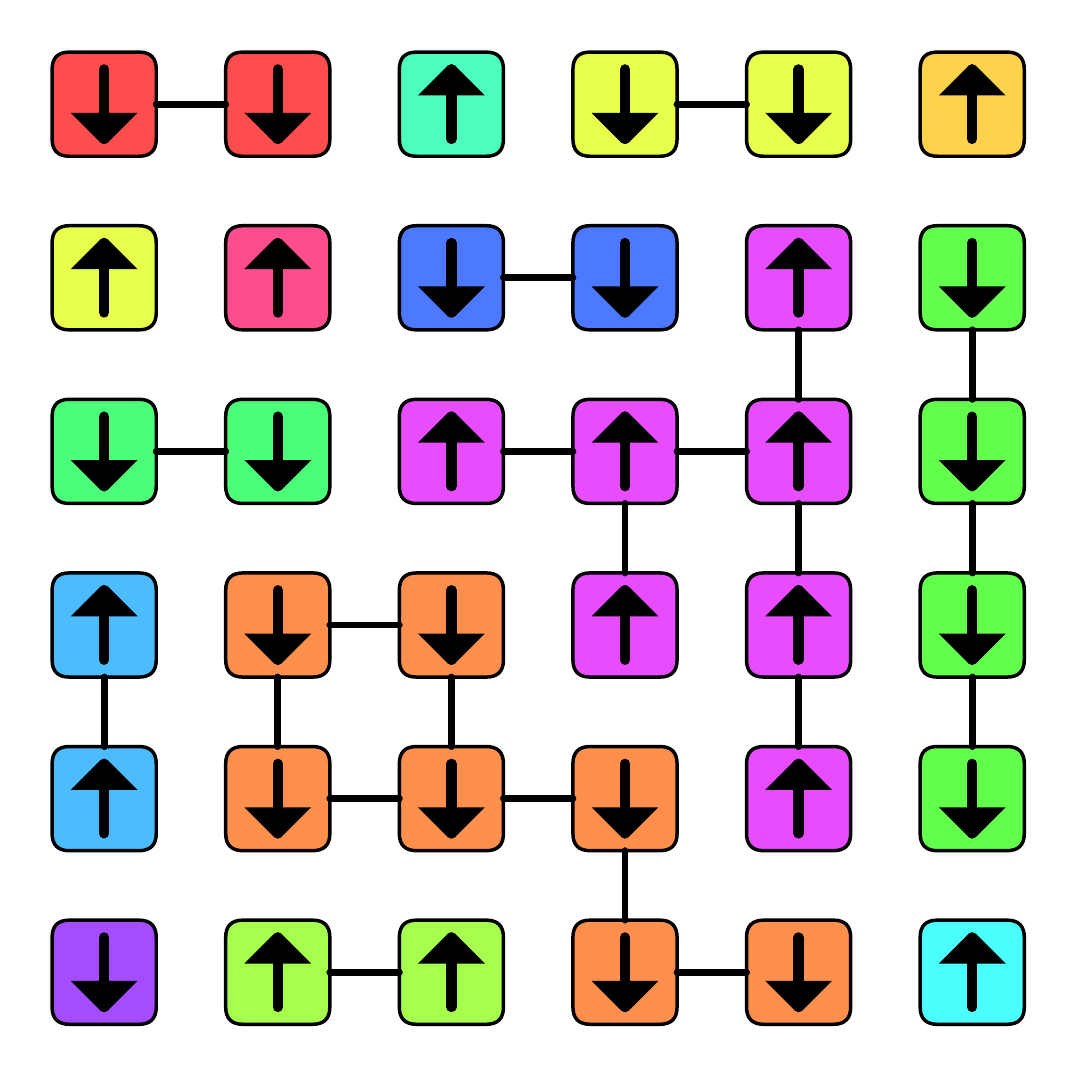}
                \caption{The formation of clusters is done using a label-equivalence algorithm designed for massively parallel systems.  It is able to converge quickly even on large lattices, and assigns each cluster a label such that all connected components share the same label.  Here the different colors are used to represent different clusters.}
                \label{fig:formingClusters}
            \end{minipage} \qquad
            \begin{minipage}{\figureWidth}
                \centering
                \textbf{Assignment of New Configuration}\par\medskip
                \includegraphics[width=0.8\linewidth]{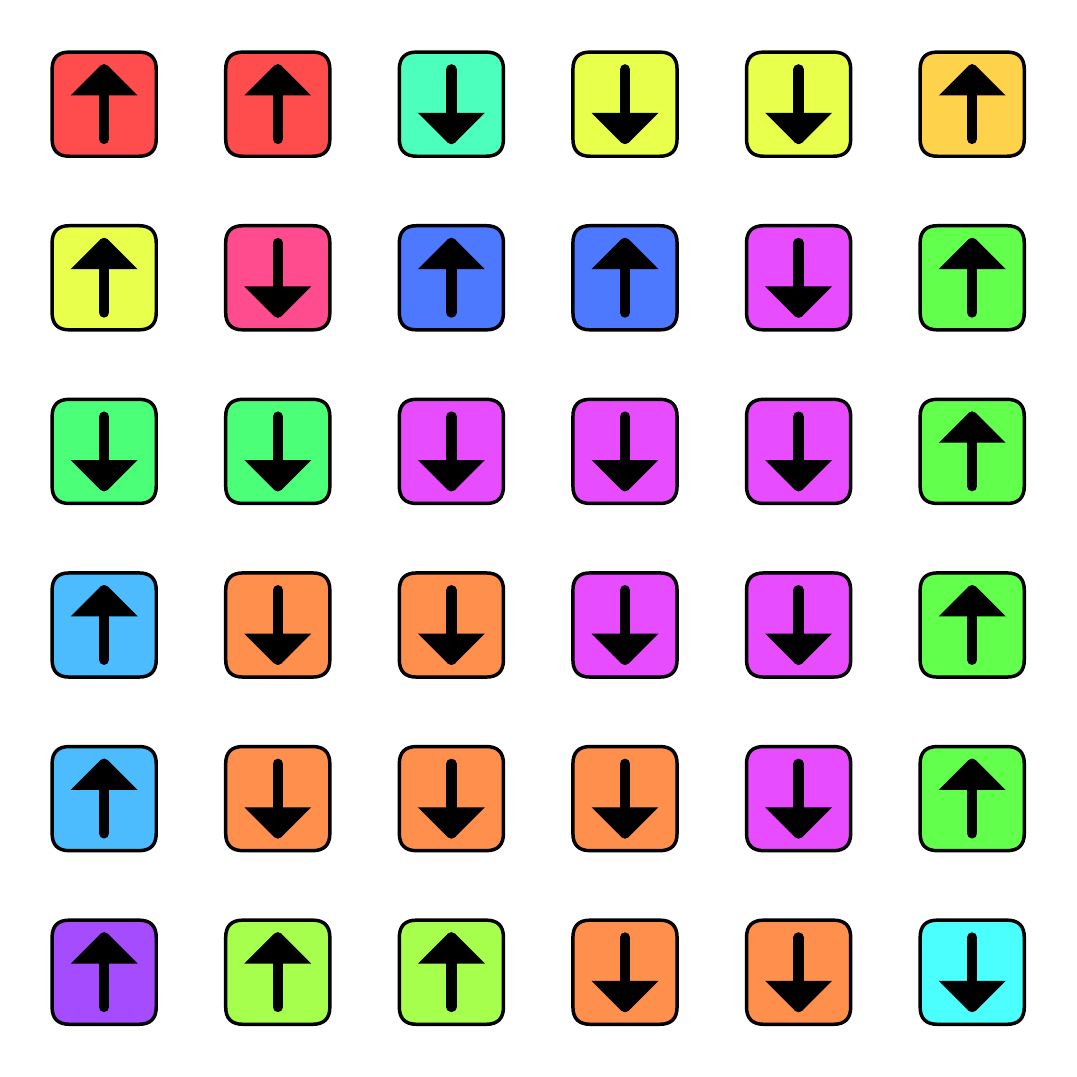}
                \caption{A new spin value is then generated for each cluster, and a thread per site sets each location's new spin value.  This creates a new arrangement of spins which can be very different from the prior configuration while maintaining the expected physical properties.}
                \label{fig:newConfig}
            \end{minipage} \qquad
        \end{figure}
        
    \section{Wolff Clustering Algorithm}
        Similarly, clustering algorithms can be applied to continuous spin models in higher dimension.  In particular, the O(4) symmetric $\phi^4$ theory in four dimensions was simulated using the Wolff clustering algorithm \cite{wolff_collective_mc}.  It was  shown by Frick et al.~\cite{phifour_clustering} that this algorithm exhibits the same desirable properties of the Swendsen-Wang algorithm; thus it is once again desirable to develop a GPU accelerated implementation.  Though the ideas are similar, there are a few critical differences which must be addressed.  
        
        Most trivially, the implementation must be extended to work in four dimensions.  This was achieved by extending the array used to hold site values to a four dimensional array.  When checking nearest neighbors to form bonds, each site now checks the next index for each of the four array elements.  The other significant change comes from the change to a continuous spin model.  No longer can the values of neighboring sites be  compared in a binary fashion.  Instead, bonds between sites are formed with the following probability, where $\kappa$ is the coupling constant and $r$ is a random four component direction normalized such that $r^\alpha r^\alpha \equiv \sum_\alpha r^\alpha r^\alpha=1$:
        \begin{equation}
            p=1-\hbox{exp}\{\hbox{min}(0,\;-4 \kappa \phi^\alpha_x r^\alpha \phi^\beta_{x+\mu} r^\beta)\}
        \end{equation}
        That is, the lattice field is projected into the direction of the four-vector $r^\alpha$ and this projection is the basis of the probability.
        Finally, we must redefine what it means to flip clusters.  Once they have been formed from the bonds created, clusters formed are selected and joined into a multicluster $C$ with probability $\frac{1}{2}$.  They are then reflected by the random direction used to form bonds by the following:
        \begin{equation}
            (R^C\phi)_x = 
            \begin{cases} 
                \phi_x - 2(\phi^\alpha_x r^\alpha)r & \mbox{if } \phi_x \in C \\
                \phi_x & \mbox{if } \phi_x \not\in C
            \end{cases}
        \end{equation}

    \section{Results}
\begin{figure}
            \centering
            \includegraphics[width=0.45\linewidth]{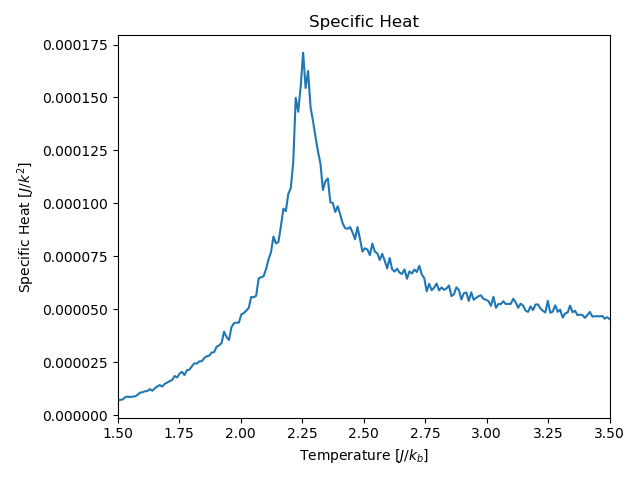}
            \caption{The specific heat per site of the Ising lattice.  The specific heat would be expected to drastically increase at the critical point, which we can clearly see happening from the peak in the neighborhood of $k_B T=2.25$. \label{Cfig}}        
        \end{figure}

        To verify the correctness of the new algorithm, the magnetization, energy, and specific heat across a range of temperatures was calculated and compared to both the CPU algorithm's results as well as previously found results \cite{ising1} \cite{ising2}. Figure \ref{Cfig} shows the specific heat for the Ising system as a function of temperature.  It shows the characteristic peak near the critical temperature where long range order disappears. 

        \begin{figure}
            \centering
            \includegraphics[width=0.45\linewidth]{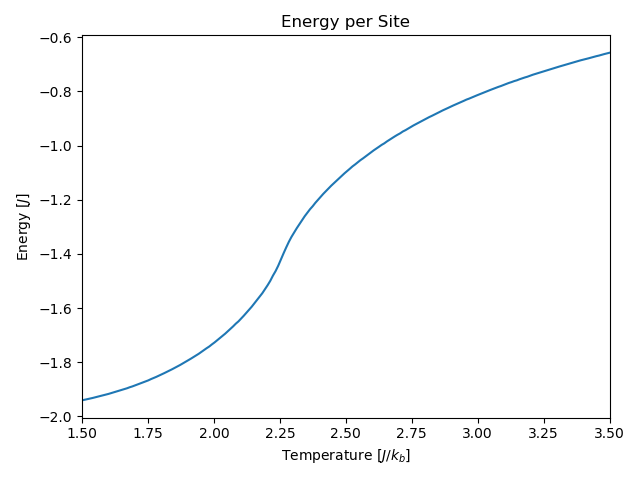}
            \caption{The energy per site of the Ising lattice as a function of the system temperature.  Notice the inflection point around $k_bT=2.25$.  This denotes a phase change, signaling that this is a critical point of the system. }
        \end{figure}

        \begin{figure}
            \centering
            \includegraphics[width=0.45\linewidth]{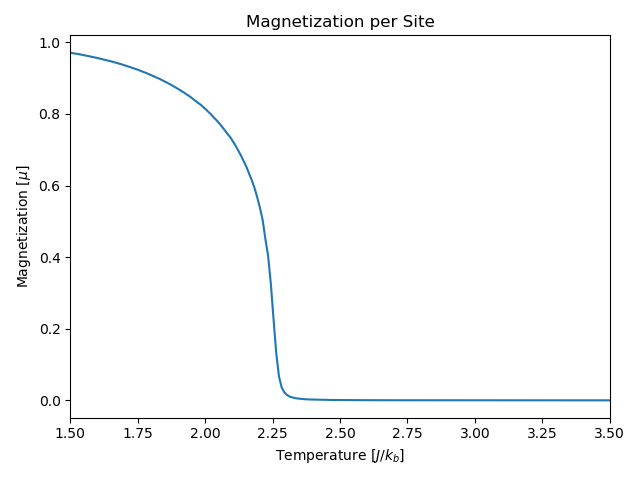}
            \caption{The magnetization per site of the Ising lattice as a function of temperature.  The critical point is once again apparent, above it the magnetization of the system goes to approximately zero.}
        \end{figure}

        All results generated by the GPU implementation of the algorithm very closely match the sets of data generated by the serial CPU algorithm running an equivalent simulation.  Additionally, the behaviors and locations of the critical points are equivalent to those described by the above resources.  This give confidence in the implementations correctness and accuracy.

    \section{Performance}
        To serve as a benchmark of the GPU accelerated codes performance, a serial algorithm running on traditional CPUs was developed.  Resources at the Pittsburgh Supercomputing Center were utilized to compare performance; the serial code was executed on a Intel Haswell E5-2695 process while the GPU code was executed on a NVIDIA Tesla P100 card.  On larger grid sizes, the parallel algorithm outperformed the serial code by about a factor of over 17 times on sampled grid sizes.   

        \begin{center}
            \begin{figure}[H]
                \centering
                \includegraphics[width=0.5\linewidth]{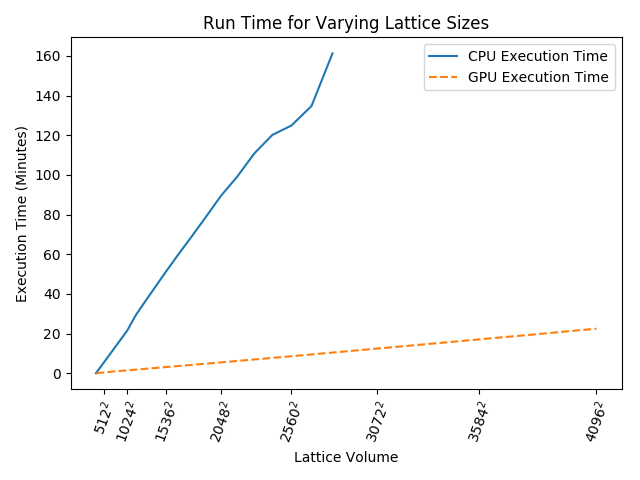}
                \caption{Equivalent simulations were executed on both CPU and GPU code.}
            \end{figure}

            {10,000 Iterations at $k_bT=2.5$}
            \begin{tabular}{|c|c|c|c|}
                \hline
                Lattice Length & CPU Time (minutes) & GPU Time (minutes) & GPU Speedup \\
                \hline
                128 & 0.339 & 0.112 & 2.944 \\
                \hline
                256 & 1.369 & 0.151 & 9.057 \\
                \hline
                512 & 5.367 & 0.390 & 13.765 \\
                \hline
                1024 & 21.497 & 1.472 & 14.604 \\
                \hline
                2048 & 89.504 & 5.483 & 16.325 \\ 
                \hline
                4096 & 400.372 & 22.440 & 17.842 \\
                \hline
            \end{tabular}
        \end{center}
        
        The performance of this algorithm proved to be bounded by the GPUs memory bandwidth.  To mitigate this problem, shared and local memory caches on the device were utilized as much as possible to avoid having slow global memory accesses.  A commonly used technique in this code is to load a local section of the lattice into shared memory at the start of kernel execution.  This allows for the required global memory accesses to happen in a coalesced fashion, fully taking advantage of the GPUs warp based architecture.  Some improvements still exist to be made in the handling of shared memory caches, particularly in the edge cases where the state from another thread block is required.  However, some of these locations, particularly those directly left and right of the current thread block cannot be efficiently moved into shared memory because they are too far apart in memory \cite{globalMemory}.  Loading a row of the thread blocks local lattice currently takes one transfer operation because its width is 32 sites, which is the size of a warp.  Loading the left and right neighbors into shared memory at the same time would take and additional memory transaction each.  Therefore, it is left to be scheduled by the compiler to be hidden by switching which warp is being executed while the require memory is accessed.
        
        \begin{center}
            \begin{figure}[H]
                \centering
                \includegraphics[width=0.5\linewidth]{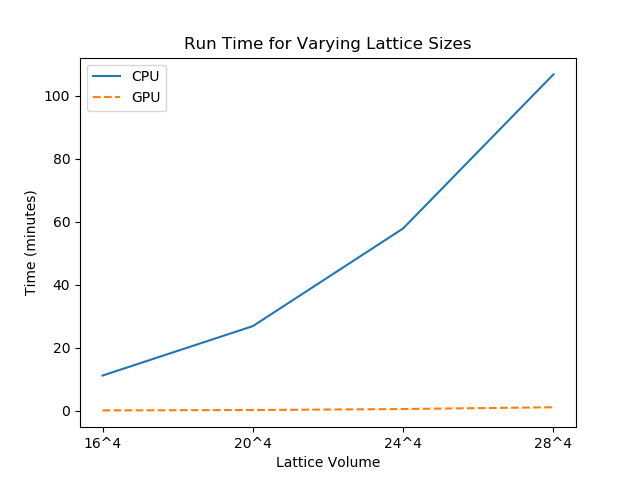}
                \caption{$\phi^4$ simulation results. Equivalent simulations were executed on both CPU and GPU code.}
            \end{figure}

            {10,000 Iterations at $\kappa=0.304$}
            \begin{tabular}{|c|c|c|c|}
                \hline
                Lattice Length & CPU Time (minutes) & GPU Time (minutes) & GPU Speedup \\
                \hline
                16 & 11.246 & 0.186 & 60.188 \\
                \hline
                20 & 26.931 & 0.305 & 88.218 \\
                \hline
                24 & 57.939 & 0.619 & 93.589 \\
                \hline
                28 & 106.821 & 1.188 & 89.907 \\
                \hline
            \end{tabular}
        \end{center}
        
        The Wolff cluster algorithm benefits even more than the Swendsen Wang algorithm from a GPU implementation. Largely, this is because there is more computation required for each step, so although the memory bandwidth is still being saturated, it is less of a limiting factor.  The additional complexity of calculating the bond probabilities and the reflection of flipped clusters give a large boost to the parallelization of a GPU.

        Dynamic parallelization was also utilized to minimize memory transfers between device and host memory.  It allows the entire clustering operation to be completed from a single kernel launched from the CPU.  This maintenance kernel manages the three phases of the clustering process and watches for the conclusion of the algorithm without continually passing a flag variable from the device to the host.

    \section{Conclusions}
        As mentioned above, the Swendsen-Wang algorithm allows simulations to occur around the critical points of systems without the large autocorrelation times experienced by other algorithms, such as the Metropolis-Hasting algorithm.  By accelerating the rate at which such system can be evolved with efficient GPU code, an even greater number of configurations can be sampled.  This combination of a cluster method and GPU accelerations allows for a better understanding the system's properties near the critical temperature.  We have shown that such acceleration is also possible in the Wolff algorithm, using similar algorithmic techniques.  Another  statistical model where such methods could be applied is the Potts model.  We have found that a speedup of one or two orders of magnitude is possible, in spite of the challenges posed by the significant latency of global memory accesses on a GPU.  This is achieved through well-designed algorithms that minimize this potential bottleneck.  \\
 
	\section{Acknowledgements}
        We would like to acknowledge the Department of Energy, Office of Science, Office of High Energy Physics for their support through Grant Number. DE-SC0013496.  Additionally, this work used the Extreme Science and Engineering Discovery Environment (XSEDE), which is supported by the National Science Foundation grant number ACI-1548562.  The Pittsburgh Supercomputing Center was utilized using the grant "Many-core accelerated lattice field theory | TG-PHY150039".  Additional funding was provided by the Rensselaer Undergraduate Research Program. 

\bibliography{references}

\end{document}